\documentclass[sn-mathphys]{sn-jnl}
\jyear{2021}%

\begin{document}

\title[Article Title]{Fundamental connection between temperature-quenched 2D superfluids and 2D quantum turbulence}

\author*[1]{\fnm{Gary A.} \sur{Williams}}\email{gaw@ucla.edu}

\affil*[1]{\orgdiv{Department of Physics \& Astronomy}, \orgname{University of California}, \orgaddress{\street{475 Portola Plaza}, \city{Los Angeles}, \postcode{90095}, \state{CA}, \country{USA}}}
 

\abstract{There is a fundamental connection between temperature-quenched 2D superfluids and 2D quantum turbulence: the mechanism responsible for the decay of the vorticity after the quench is the enstrophy cascade of 2D turbulence.  The range of the cascade is shown to increase with time after the quench as $t^{1/2}$, being equal to the dynamic scaling length characterizing the quench.  These results may apply to other phase-ordering transitions involving the decay of topological objects.}


\maketitle
\section{Introduction}

The static equilibrium properties of the two-dimensional (2D) superfluid phase transition are now well understood in terms of the Kosterlitz-Thouless theory \cite{kosterlitzrev} involving thermally excited vortex-antivortex pairs.  However, the nonequilibrium dynamics of such a system is less well characterized.   An example of this is a sudden temperature quench of the superfluid that leaves a non-equilibrium high density of vortices at the new low temperature, and the only way they can decay is by vortex-antivortex annihilation, but this takes time due to the friction on the normal vortex cores.  Early studies of the vortex dynamics by Minnhagen and co-workers  \cite{minn2z} in 2D XY model simulations found puzzling results, that there seemed to be be two different dynamic exponents, depending on the boundary conditions used in the simulations.  One dynamic exponent found only with fluctuating-twist boundary conditions took the form $z_{scale}  = 2\pi K  - 2$, where $K = \hbar^2\sigma _s (T )/(m^2 k_B T)$ with $\sigma _s$ the macroscopic 2D superfluid areal density and $m$ the atomic mass.  This exponent varies with the temperature $T$, taking a value $z_{scale}  = 2$ at the critical temperature $T_{KT}$ where $K = 2/\pi$, and then increasing at temperatures below $T_{KT}$.  A second dynamic exponent found with all boundary conditions was the constant value $z \approx 2$.  The presence of two different dynamic exponents was unusual, and it was not entirely clear how these came into the vortex dynamics.  An exact solution \cite{forrester2013} for temperature quenches starting from $T_{KT}$ and below shed light on some of these questions, where the vortex density was found to decay with time $t$ after the quench as $t^{-z_{scale} / z}$.  Here $z_{scale}$ is evaluated at the initial temperature before the quench, and $z = 2$ is exact.
 
Another nonequilibrium situation in 2D superfluids is the case of quantum turbulence, which is the case when extra vortices are injected into the superfluid.  By analogy with turbulence results in 2D classical fluids \cite{kraichnan80,boffeta12}, it was expected that the turbulence in superfluids would self-organize into constant-flux cascades of both energy and vorticity (enstrophy).  This was indeed observed in computer simulations, where a cascade with a constant flux of energy from small to large scales was found with a Kolmogorov $k^{-5/3}$ energy spectrum \cite{reevesdirect} .  A constant-flux enstrophy cascade from large scales to small was also observed in a different simulation, with the expected $k^{-3}$ energy spectrum \cite{reeves}, as had been predicted much earlier \cite{turb2001}.  Analytic solutions have been found for the enstrophy cascade \cite{turbprf}, which is a forward cascade where vortex pairs of large separation are injected into the superfluid at a constant rate, and due to the friction on the vortex cores they diffuse at a constant flux to smaller separations, finally annihilating at the vortex core scale at the same rate they are being injected. 

We have found that these seemingly unrelated nonequilibrium states have a fundamental connection: the mechanism of the decay of the vorticity after the temperature quench is the enstrophy cascade.  There is a constant flux of vortices from the large separation scales present at the initial temperature, moving to the small scale of the vortex core size where they annihilate, to finally get to the very low vortex density of the final temperature.  This is precisely the definition of the enstrophy cascade, and the range of the cascade is found to grow as the dynamic length, varying as $t^{1/z}$.

\section{Temperature quenches}  
\begin{figure}[t]
\begin{center}
\includegraphics[width=0.9\textwidth]{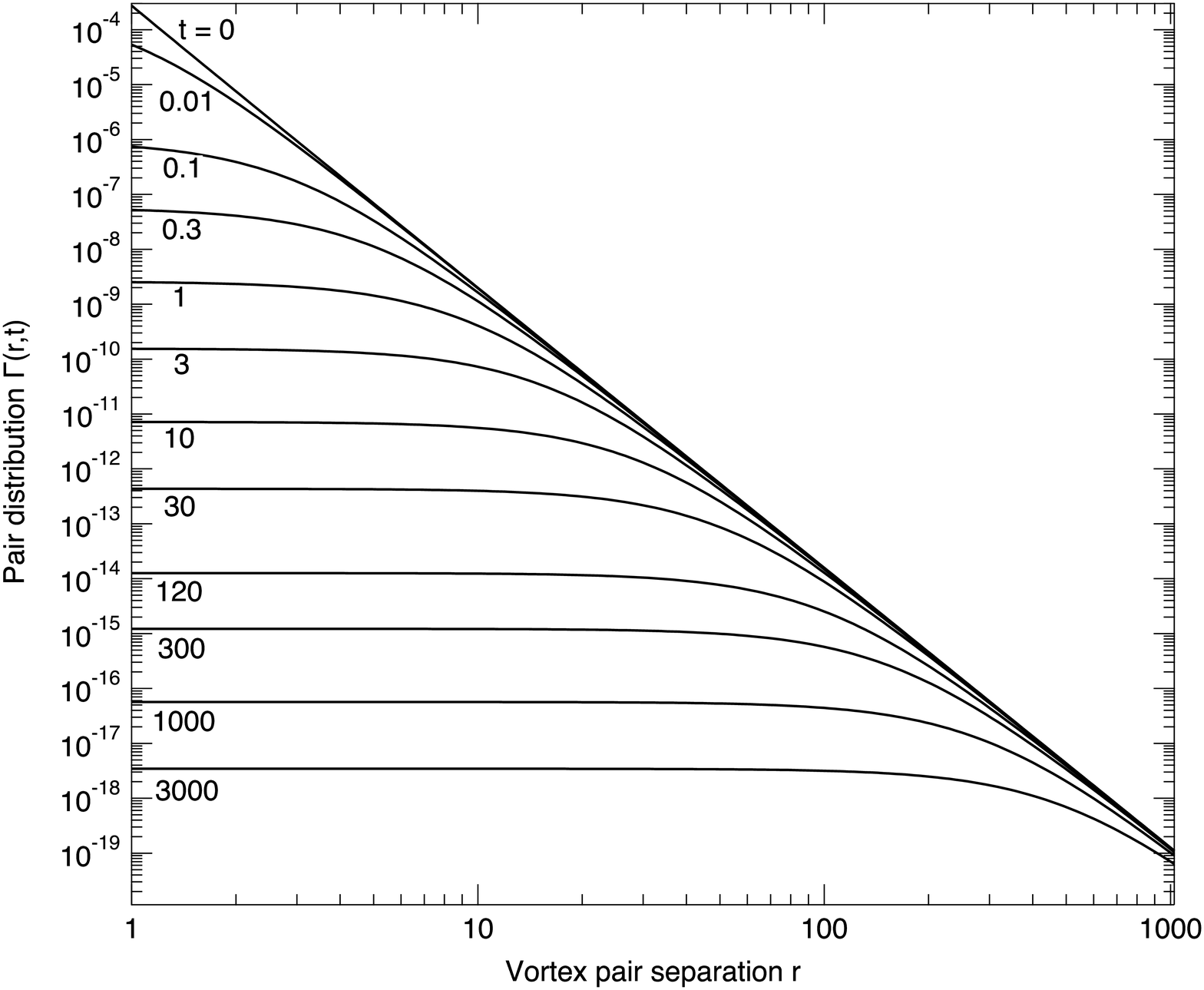}
\end{center}
\caption{Vortex pair distribution function $\Gamma$ as a function of the pair separation $r$ (in units of $a_0$) and time $t$ (in units of the diffusion time).  }
\label{fig1}\end{figure}

To characterize an instantaneous temperature quench from an initial temperature $T_i$ to a final low temperature $T_f$ it is necessary to solve a Fokker-Planck equation \cite{ahns} for the distribution $\Gamma (r,t)$ of vortex-antivortex pairs of separation $r$ at time $t$,
\begin{equation}
\frac{{\partial \,\Gamma}}{{\partial \,t}} = \frac{1}{r}\,\;\frac{\partial }{{\partial r}}\left( {r\frac{{\partial \Gamma  }}{{\,\partial r}} + 2\pi K\,\Gamma  } \right)
\label{eq1}
\end{equation}
where this is made dimensionless by taking $r$ in units of the vortex core size $a_0$, $\Gamma$ in units $a_0^{-4}$, and $t$ in units of the diffusion time $\tau_0 = a_0^2 / 2D$ with $D$ the vortex diffusion coefficient at the temperature $T_f$ set by the thermal bath (which e.g.\, would be the substrate of a thin superfluid helium film).  The vortex motion is diffusive even at low temperatures, as shown by experiments on helium films below 1 K \cite{agnolet,emin,Hieda}.

$K$ in Eq.\,\ref{eq1} is determined by simultaneously solving the Kosterlitz recursion relation \cite{kosterlitz} for the superfluid density
\begin{equation}
\frac{{\partial K}}{{\partial r}} =  - 4\pi ^3 r^3 K^2 \Gamma \quad . 
\label{eq2}   
\end{equation}
For quenches starting from below $T_{KT}$ we can make the approximation that $K_i$ is effectively a constant, since the spatial renormalization from Eq.\,(\ref{eq2}) is rapid, changing from the initial $K_{0i}$ to the renormalized $K_i$ over a length scale that can be less than a core radius.  In that limit the initial distribution at $t = 0$ is given by $\Gamma_0 ' \,r^{-2 \pi K_i}$ with $\Gamma_0 '$ a constant depending on the vortex core energy.  The solution for the pair distribution after the quench where now $K = K_f  \simeq const $ is given by 
\begin{equation}
\Gamma (r,t) = \beta \,\,_1 F_1 \left[ {\pi K_i ,1 + \pi K_f , - \frac{{r^2 }}{{2z{\kern 1pt} \,t^{2/z} }}} \right]\;\,t^{ - 2\pi K_i /z} 
\label{eq3}
\end{equation}
where $_1 F_1$ is the confluent hypergeometric function of the first kind, $z$ = 2 exactly, and $\beta$ is a constant depending on $K_i$ and $K_f$.  Figure\,\ref{fig1} shows a plot of $\Gamma (r,t)$ for a quench from 0.9 $T_{KT}$ to 0.1 $T_{KT}$, where $K_i = 0.814$ and  $K_f  = 7.479$.  Just after the quench only the smallest-separation vortices start to disappear, since they get to the core-size annihilation scale $r = 1$ the quickest, while the distribution at larger scales remains at the equilibrium variation $r^{-2 \pi K_i} = r^{-5.11}$.  Note in particular at later times that the distribution develops a very flat region at shorter separations, and that the length scale of the flat regions increases with increasing time.

The density of vortices can then be found by integrating over $r$,
\begin{equation}
\rho_v (t) = 2 \int_1^\infty  {\Gamma (r,t)\,2\pi r} dr = \beta' \,\, _1 F_1 [ {\pi K_i  - 1,\:\pi K_f ,\: - \frac{{t^{ - 2/z} }}{{2z}}} ]\,\, t^{ - z_{scale} /z} 
\label{eq4}
\end{equation}
where $\beta'$ is a constant, and $z_{scale} = 2 \pi K_i - 2$.  At times longer than a few diffusion times the hypergeometric function goes to a constant, and the vortex decay is then  $t^{ - z_{scale} /z} = t^{-1.557}$.   The exponent $z_{scale}$ appears in the dynamics due to the second term on the right-hand side of Eq.\,1, which reflects the attractive interaction between the vortices of the pair, while the factor of $z = 2$ appears because of the second derivative of the first term in Eq.\,1, which gives rise to the diffusive motion of the vortices.  This confirms the result of Minnhagen et al. \cite{minn2z} that two dynamic exponents control the dynamics of the superfluid, but it is still not entirely clear why the boundary conditions play a role in this.  The factor $2 \pi K_i $ appears in $z_{scale}$ because the initial distribution function $\Gamma_i$ is proportional to $r^{-2 \pi K_i}$, and this gives a very simple interpretation of why the vortex decay becomes more rapid at lower temperatures where $K_i$ increases:  the steeper distribution means most of the pairs are closer together, and hence annihilate more rapidly.  This has nothing to do with any Kibble-Zurek "creation" of vortices  in the quench \cite{zurek1985,zurekkt}.

\section{Enstrophy cascade}
\begin{figure}[t]
\begin{center}
\includegraphics[width=0.9\textwidth]{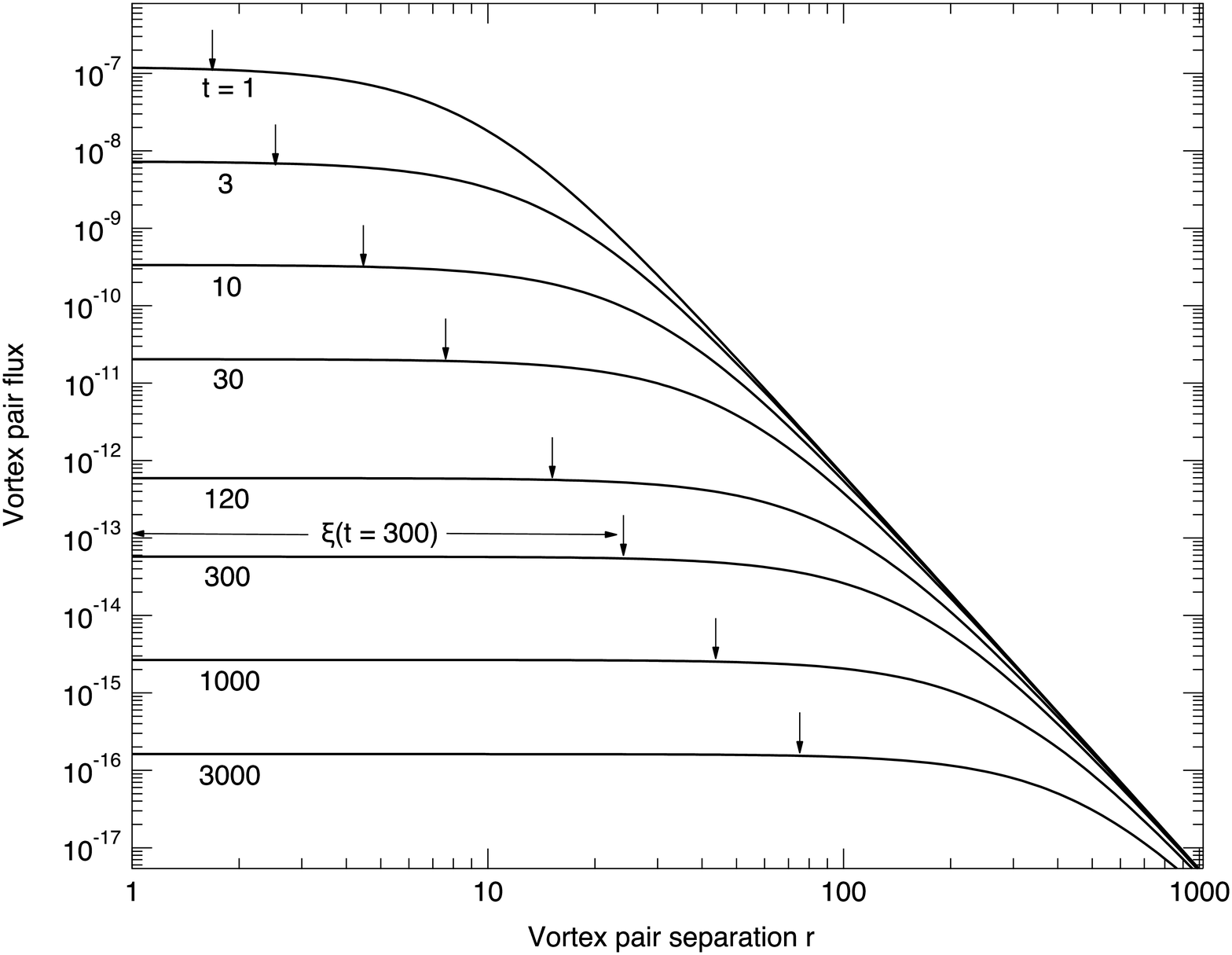}
\end{center}
\caption{Vortex pair flux computed from the distributions of Fig\,.1 as a function of the pair separation $r$ (in units of $a_0$) and time $t$ (in units of the diffusion time).  }
\label{fig2}\end{figure}

Analytic solutions for the enstrophy cascade can be found \cite{turbprf} using the Fokker-Planck Eq.\,1, but with a 2D Dirac delta function added to the right-hand side to inject vortex pairs of large separation $R$.  The injection rate multiplying the delta function in dimensionless form is $\alpha =  a_0^2\dot Q\tau _0$ where $\dot Q$ is the number of vortex pairs of separation $R$ injected per unit area per time.  The thermal bath is held at a very low temperature (i.e. 0.1 $T_{KT}$) so that no thermal vorticity is present.  Due to the contact with the thermal bath, negative-temperature Onsager states of the vortices \cite{simula} cannot form, and hence the energy cascade will not be present.  For the comparison with quenched superfluids we only consider the case where $\alpha$ is small enough that the resulting vortex density is well below that at the Kosteritz-Thouless transition.  In that limit the superfluid density is unaffected by the vortices, so $K = K_0 = const$, the "bare" superfluid density, and the solution for the distribution function characterizing the enstrophy cascade is
\begin{equation}
\begin{array}{l}
\Gamma (r)  = {\Gamma _0} = \alpha /2\pi {K_0}\quad \quad (r < R)\\
\;\;\;\;\;\;\;\, = {\Gamma _0}{\left( {r/R} \right)^{ - 2\pi {K_0}}}\quad\quad\;\, (r > R).
\end{array}
\label{eq5}
\end{equation}  
For $r > R$ the solution is a quasi-thermal distribution extending from $R$, which arises from injected pairs initially at separation $R$ getting a thermal kick to higher separation.  Note for comparison with the quench case that the pair distribution of the cascade is completely constant for the cascade range $r < R$.  This is a solution with a constant flux of vortices from the large scale $R$ to the small core size scale, since the flux is
 $ {r{\partial \Gamma } \mathord{\left/
 {\vphantom {{\partial \Gamma } {\partial r + 2\pi K \Gamma}}} \right.
 \kern-\nulldelimiterspace} {\partial r + 2\pi K \Gamma}} = \alpha = const$.  
 
 The energy spectrum corresponding to Eq.\,\ref{eq5} varies as $k^{-3}$ \cite{turbprf}, similar to the classical enstrophy spectrum \cite{kraichnan80,boffeta12}, though it is not quite the same.  It is proportional to $\eta \tau_0$ where $\eta$ is the k-space enstrophy injection rate and $\tau_0$ the diffusion time.  The classical spectrum is proportional to $\eta^{2/3}$ since the vorticity there has all different diffusion characteristics, whereas the quantum vortices all have the same $\tau_0$.
 
\begin{figure}[t]
\begin{center}
\includegraphics[width=0.9\textwidth]{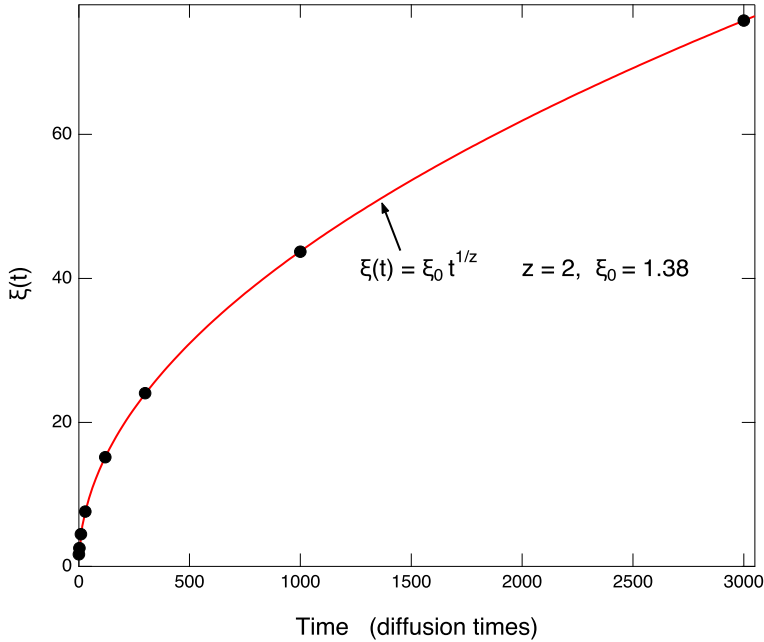}
\end{center}
\caption{Dynamic lengths (arrow positions) from Fig.\,2 versus time. }
\label{fig3}\end{figure}

 For further comparison with the quench case we can compute the flux from the quench distributions of Fig.\,\ref{fig1},
 shown in Figure\,\ref{fig2}. The flux is seen to be nearly constant over separations that increase with time, before falling off at large scales.  We can arbitrarily define a dynamic length $\xi$ as the length where the flux has decreased by 5\% from its value at $r = 1$, marked by the arrows in Fig.\,\ref{fig2}.  Figure\,\ref{fig3} shows the dynamic lengths as a function of time, and is completely well fit by $\xi (t) = {\xi _0}\,{t^{1/2}}$ with ${\xi _0} = 1.38$, the known dynamic length associated with $z = 2$.  This is strong proof that the mechanism of the vortex decay after the quench is an enstrophy cascade, with the range of the constant-flux cascade given by the increasing dynamic length.

\section{Conclusions} 

We think it is likely that the connection between turbulent cascades and phase-ordering dynamics may apply more generally.  There are numerous systems with phase-ordering transitions that involve topological excitations \cite{bray}, and it would be interesting to look for constant-flux cascades in these systems.  One well-known example is quenches of liquid crystals from the isotropic to nematic state \cite{yeomans,puri}, where disclinations in the director field form annihilating topological objects.  

A further note is that the 2D superfluid system considered here is actually not in the same dynamic universality class as the 2D XY model, even though they are in the same 2D O(2) static class.  Dissipation in a superfluid vortex is entirely localized to the normal vortex core, while dissipation in XY vortices occurs over finite distances from the core \cite{radz}.  This leads to a scale-dependent mobility of the XY vortices \cite{mobility}, and gives rise to a log correction term in the dynamic scaling \cite{yurke}.


\begin{thebibliography}{23}
\ifx \bisbn   \undefined \def \bisbn  #1{ISBN #1}\fi
\ifx \binits  \undefined \def \binits#1{#1}\fi
\ifx \bauthor  \undefined \def \bauthor#1{#1}\fi
\ifx \batitle  \undefined \def \batitle#1{#1}\fi
\ifx \bjtitle  \undefined \def \bjtitle#1{#1}\fi
\ifx \bvolume  \undefined \def \bvolume#1{\textbf{#1}}\fi
\ifx \byear  \undefined \def \byear#1{#1}\fi
\ifx \bissue  \undefined \def \bissue#1{#1}\fi
\ifx \bfpage  \undefined \def \bfpage#1{#1}\fi
\ifx \blpage  \undefined \def \blpage #1{#1}\fi
\ifx \burl  \undefined \def \burl#1{\textsf{#1}}\fi
\ifx \doiurl  \undefined \def \doiurl#1{\url{https://doi.org/#1}}\fi
\ifx \betal  \undefined \def \betal{\textit{et al.}}\fi
\ifx \binstitute  \undefined \def \binstitute#1{#1}\fi
\ifx \binstitutionaled  \undefined \def \binstitutionaled#1{#1}\fi
\ifx \bctitle  \undefined \def \bctitle#1{#1}\fi
\ifx \beditor  \undefined \def \beditor#1{#1}\fi
\ifx \bpublisher  \undefined \def \bpublisher#1{#1}\fi
\ifx \bbtitle  \undefined \def \bbtitle#1{#1}\fi
\ifx \bedition  \undefined \def \bedition#1{#1}\fi
\ifx \bseriesno  \undefined \def \bseriesno#1{#1}\fi
\ifx \blocation  \undefined \def \blocation#1{#1}\fi
\ifx \bsertitle  \undefined \def \bsertitle#1{#1}\fi
\ifx \bsnm \undefined \def \bsnm#1{#1}\fi
\ifx \bsuffix \undefined \def \bsuffix#1{#1}\fi
\ifx \bparticle \undefined \def \bparticle#1{#1}\fi
\ifx \barticle \undefined \def \barticle#1{#1}\fi
\bibcommenthead
\ifx \bconfdate \undefined \def \bconfdate #1{#1}\fi
\ifx \botherref \undefined \def \botherref #1{#1}\fi
\ifx \url \undefined \def \url#1{\textsf{#1}}\fi
\ifx \bchapter \undefined \def \bchapter#1{#1}\fi
\ifx \bbook \undefined \def \bbook#1{#1}\fi
\ifx \bcomment \undefined \def \bcomment#1{#1}\fi
\ifx \oauthor \undefined \def \oauthor#1{#1}\fi
\ifx \citeauthoryear \undefined \def \citeauthoryear#1{#1}\fi
\ifx \endbibitem  \undefined \def \endbibitem {}\fi
\ifx \bconflocation  \undefined \def \bconflocation#1{#1}\fi
\ifx \arxivurl  \undefined \def \arxivurl#1{\textsf{#1}}\fi
\csname PreBibitemsHook\endcsname

\bibitem{kosterlitzrev}
\begin{barticle}
\bauthor{\bsnm{Kosterlitz}, \binits{J.M.}}:
\batitle{Kosterlitz--Thouless physics: a review of key issues}.
\bjtitle{Rep. Prog. Phys.}
\bvolume{79},
\bfpage{026001}
(\byear{2016})
\doiurl{10.1088/0034-4885/79/2/026001}
\end{barticle}
\endbibitem

\bibitem{minn2z}
\begin{barticle}
\bauthor{\bsnm{Jensen}, \binits{L.M.}},
\bauthor{\bsnm{Kim}, \binits{B.J.}},
\bauthor{\bsnm{Minnhagen}, \binits{P.}}:
\batitle{Dynamic critical exponent of two-, three-, and four-dimensional
  $\mathrm{XY}$ models with relaxational and resistively shunted junction
  dynamics}.
\bjtitle{Phys. Rev. B}
\bvolume{61},
\bfpage{15412}
(\byear{2000})
\doiurl{10.1103/PhysRevLett.87.037002}
\end{barticle}
\endbibitem

\bibitem{forrester2013}
\begin{barticle}
\bauthor{\bsnm{Forrester}, \binits{A.}},
\bauthor{\bsnm{Chu}, \binits{H.-C.}},
\bauthor{\bsnm{Williams}, \binits{G.A.}}:
\batitle{Exact solution for vortex dynamics in temperature quenches of
  two-dimensional superfluids}.
\bjtitle{Phys. Rev. Lett.}
\bvolume{110},
\bfpage{165303}
\doiurl{10.1103/PhysRevLett.110.165303}
(\byear{2013})
\end{barticle}
\endbibitem

\bibitem{kraichnan80}
\begin{barticle}
\bauthor{\bsnm{Kraichnan}, \binits{R.H.}},
\bauthor{\bsnm{Montgomery}, \binits{D.}}:
\batitle{Two-dimensional turbulence}.
\bjtitle{Reports on Prog. Phys.}
\bvolume{43}(\bissue{5}),
\bfpage{547}
(\byear{1980})
\end{barticle}
\endbibitem

\bibitem{boffeta12}
\begin{barticle}
\bauthor{\bsnm{Boffetta}, \binits{G.}},
\bauthor{\bsnm{Ecke}, \binits{R.E.}}:
\batitle{Two-dimensional turbulence}.
\bjtitle{Ann. Rev. Fluid Mech.}
\bvolume{44},
\bfpage{427}
(\byear{2012})
\doiurl{10.1146/annurev-fluid-120710-101240}
\end{barticle}
\endbibitem

\bibitem{reevesdirect}
\begin{barticle}
\bauthor{\bsnm{Reeves}, \binits{M.T.}},
\bauthor{\bsnm{Billam}, \binits{T.P.}},
\bauthor{\bsnm{Anderson}, \binits{B.P.}},
\bauthor{\bsnm{Bradley}, \binits{A.S.}}:
\batitle{Inverse energy cascade in forced two-dimensional quantum turbulence}.
\bjtitle{Phys. Rev. Lett.}
\bvolume{110},
\bfpage{104501}
(\byear{2013}).
\doiurl{10.1103/PhysRevLett.110.104501}
\end{barticle}
\endbibitem

\bibitem{reeves}
\begin{barticle}
\bauthor{\bsnm{Reeves}, \binits{M.T.}},
\bauthor{\bsnm{Billam}, \binits{T.P.}},
\bauthor{\bsnm{Yu}, \binits{X.}},
\bauthor{\bsnm{Bradley}, \binits{A.S.}}:
\batitle{Enstrophy cascade in decaying two-dimensional quantum turbulence}.
\bjtitle{Phys. Rev. Lett.}
\bvolume{119},
\bfpage{184502}
(\byear{2017}).
\doiurl{10.1103/PhysRevLett.119.184502}
\end{barticle}
\endbibitem

\bibitem{turb2001}
\begin{bchapter}
\bauthor{\bsnm{Chu}, \binits{H.-C.}},
\bauthor{\bsnm{Williams}, \binits{G.A.}}:
\bctitle{Nonequilibrium vortex dynamics in superfluid phase transitions and
  superfluid turbulence}.
In: \beditor{\bsnm{Barenghi}, \binits{C.F.}},
\beditor{\bsnm{Donnelly}, \binits{R.J.}},
\beditor{\bsnm{Vinen}, \binits{W.F.}} (eds.)
\bbtitle{Quantized Vortex Dynamics and Superfluid Turbulence}.
\bsertitle{Lecture Notes in Physics},
vol. \bseriesno{571},
pp. \bfpage{226}.
\bpublisher{Springer},
\blocation{Heidelberg}
(\byear{2001})
\end{bchapter}
\endbibitem

\bibitem{turbprf}
\begin{barticle}
\bauthor{\bsnm{Forrester}, \binits{A.}},
\bauthor{\bsnm{Chu}, \binits{H.-C.}},
\bauthor{\bsnm{Williams}, \binits{G.A.}}:
\batitle{Renormalized analytic solution for the enstrophy cascade in
  two-dimensional quantum turbulence}.
\bjtitle{Phys. Rev. Fluids}
\bvolume{5},
\bfpage{072701}
(\byear{2020}).
\doiurl{10.1103/PhysRevFluids.5.072701}
\end{barticle}
\endbibitem

\bibitem{ahns}
\begin{barticle}
\bauthor{\bsnm{Ambegaokar}, \binits{V.}},
\bauthor{\bsnm{Halperin}, \binits{B.I.}},
\bauthor{\bsnm{Nelson}, \binits{D.R.}},
\bauthor{\bsnm{Siggia}, \binits{E.D.}}:
\batitle{Dynamics of superfluid films}.
\bjtitle{Phys. Rev. B}
\bvolume{21},
\bfpage{1806}
(\byear{1980}).
\doiurl{10.1103/PhysRevB.21.1806}
\end{barticle}
\endbibitem

\bibitem{agnolet}
\begin{barticle}
\bauthor{\bsnm{Agnolet}, \binits{G.}},
\bauthor{\bsnm{McQueeney}, \binits{D.F.}},
\bauthor{\bsnm{Reppy}, \binits{J.D.}}:
\batitle{Kosterlitz-Thouless transition in helium films}.
\bjtitle{Phys. Rev. B}
\bvolume{39},
\bfpage{8934}--\blpage{8958}
(\byear{1989})
\doiurl{10.1103/PhysRevB.39.8934}
\end{barticle}
\endbibitem

\bibitem{emin}
\begin{barticle}
\bauthor{\bsnm{Menachekanian}, \binits{E.}},
\bauthor{\bsnm{Iaia}, \binits{V.}},
\bauthor{\bsnm{Fan}, \binits{M.}},
\bauthor{\bsnm{Chen}, \binits{J.}},
\bauthor{\bsnm{Hu}, \binits{C.}},
\bauthor{\bsnm{Mittal}, \binits{V.}},
\bauthor{\bsnm{Liu}, \binits{G.}},
\bauthor{\bsnm{Reyes}, \binits{R.}},
\bauthor{\bsnm{Wen}, \binits{F.}},
\bauthor{\bsnm{Williams}, \binits{G.A.}}:
\batitle{Superfluid onset and compressibility of $^{4}\mathrm{He}$ films
  adsorbed on carbon nanotubes}.
\bjtitle{Phys. Rev. B}
\bvolume{99},
\bfpage{064503}
(\byear{2019}).
\doiurl{10.1103/PhysRevB.99.064503}
\end{barticle}
\endbibitem

\bibitem{Hieda}
\begin{barticle}
\bauthor{\bsnm{Hieda}, \binits{M.}},
\bauthor{\bsnm{Matsuda}, \binits{K.}},
\bauthor{\bsnm{Kato}, \binits{T.}},
\bauthor{\bsnm{Matsushita}, \binits{T.}},
\bauthor{\bsnm{Wada}, \binits{N.}}:
\batitle{Extremely high frequency dependence of two-dimensional superfluid
  onset}.
\bjtitle{J. Phys. Soc. Japan}
\bvolume{78},
\bfpage{033604}
(\byear{2009}).
{\href{https://arxiv.org/abs/https://doi.org/10.1143/JPSJ.78.033604}{{https://doi.org/10.1143/JPSJ.78.033604}}}.
\doiurl{10.1143/JPSJ.78.033604}
\end{barticle}
\endbibitem

\bibitem{kosterlitz}
\begin{barticle}
\bauthor{\bsnm{Kosterlitz}, \binits{J.M.}}:
\batitle{The critical properties of the two-dimensional $\mathrm{XY}$ model}.
\bjtitle{J. Phys. C}
\bvolume{7},
\bfpage{1046}
(\byear{1974})
\end{barticle}
\endbibitem

\bibitem{zurek1985}
\begin{barticle}
\bauthor{\bsnm{Zurek}, \binits{W.H.}}:
\batitle{Cosmological experiments in superfluid helium?}
\bjtitle{Nature}
\bvolume{317}(\bissue{6037}),
\bfpage{505}
(\byear{1985}).
\doiurl{10.1038/317505a0}
\end{barticle}
\endbibitem

\bibitem{zurekkt}
\begin{barticle}
\bauthor{\bsnm{Gardas}, \binits{B.}},
\bauthor{\bsnm{Dziarmaga}, \binits{J.}},
\bauthor{\bsnm{Zurek}, \binits{W.H.}}:
\batitle{Dynamics of the quantum phase transition in the one-dimensional
  bose-hubbard model: Excitations and correlations induced by a quench}.
\bjtitle{Phys. Rev. B}
\bvolume{95},
\bfpage{104306}
(\byear{2017}).
\doiurl{10.1103/PhysRevB.95.104306}
\end{barticle}
\endbibitem

\bibitem{simula}
\begin{barticle}
\bauthor{\bsnm{Simula}, \binits{T.}},
\bauthor{\bsnm{Davis}, \binits{M.J.}},
\bauthor{\bsnm{Helmerson}, \binits{K.}}:
\batitle{Emergence of order from turbulence in an isolated planar superfluid}.
\bjtitle{Phys. Rev. Lett.}
\bvolume{113},
\bfpage{165302}
(\byear{2014}).
\doiurl{10.1103/PhysRevLett.113.165302}
\end{barticle}
\endbibitem

\bibitem{bray}
\begin{botherref}
\oauthor{\bsnm{Bray}, \binits{A.J.}}:
\batitle{Theory of phase-ordering kinetics}.
\bjtitle{Adv. Phys.}
\bvolume{357},
(\byear{1994}).
\end{botherref}
\endbibitem

\bibitem{yeomans}
\begin{barticle}
\bauthor{\bsnm{Denniston}, \binits{C.}},
\bauthor{\bsnm{Orlandini}, \binits{E.}},
\bauthor{\bsnm{Yeomans}, \binits{J.M.}}:
\batitle{Phase ordering in nematic liquid crystals}.
\bjtitle{Phys. Rev. E}
\bvolume{64},
\bfpage{021701}
(\byear{2001}).
\doiurl{10.1103/PhysRevE.64.021701}
\end{barticle}
\endbibitem

\bibitem{puri}
\begin{barticle}
\bauthor{\bsnm{Singh}, \binits{A.}},
\bauthor{\bsnm{Ahmad}, \binits{S.}},
\bauthor{\bsnm{Puri}, \binits{S.}},
\bauthor{\bsnm{Singh}, \binits{S.}}:
\batitle{Ordering dynamics of nematic liquid crystals: Monte Carlo
  simulations}.
\bjtitle{Europhys. Lett.} 
\bvolume{100},
\bfpage{36004}
(\byear{2012}).
\doiurl{10.1209/0295-5075/100/36004}
\end{barticle}
\endbibitem


\bibitem{radz}
\begin{barticle}
\bauthor{\bsnm{Radzihovsky}, \binits{L.}}:
\batitle{Anomalous energetics and dynamics of moving vortices}.
\bjtitle{Phys. Rev. Lett.}
\bvolume{115},
\bfpage{247801}
(\byear{2015}).
\doiurl{10.1103/PhysRevLett.115.247801}
\end{barticle}
\endbibitem

\bibitem{mobility}
\begin{barticle}
\bauthor{\bsnm{Dubois-violette}, \binits{E.}},
\bauthor{\bsnm{Guazzelli}, \binits{E.}},
\bauthor{\bsnm{Prost}, \binits{J.}}:
\batitle{Dislocation motion in layered structures}.
\bjtitle{Phil. Mag.\, A}
\bvolume{48},
\bfpage{727}
(\byear{1983}).
\doiurl{10.1080/01418618308236540}
\end{barticle}
\endbibitem

\bibitem{yurke}
\begin{barticle}
\bauthor{\bsnm{Yurke}, \binits{B.}},
\bauthor{\bsnm{Pargellis}, \binits{A.N.}},
\bauthor{\bsnm{Kovacs}, \binits{T.}},
\bauthor{\bsnm{Huse}, \binits{D.A.}}:
\batitle{Coarsening dynamics of the $\mathrm{XY}$ model}.
\bjtitle{Phys. Rev. E}
\bvolume{47},
\bfpage{1525}
(\byear{1993}).
\doiurl{10.1103/PhysRevE.47.1525}
\end{barticle}
\endbibitem

\end{thebibliography}
\end{document}